\documentclass[11pt]{article}
\usepackage{graphicx}
\begin{document}
~~~\\
\begin{center}
{\bf Classical probability waves  \\[1cm]}
{ Marius Grigorescu \\[3cm]  }
\end{center}
\noindent
$\underline{~~~~~~~~~~~~~~~~~~~~~~~~~~~~~~~~~~~~~~~~~~~~~~~~~~~~~~~~
~~~~~~~~~~~~~~~~~~~~~~~~~~~~~~~~~~~~~~~~~~~}$ \\[.3cm]
Probability waves in the configuration space are associated with coherent solutions of the classical Liouville or Fokker-Planck equations. Distributions localized in the momentum space provide action waves, specified by the probability density and the generating function of the Hamilton-Jacobi theory. It is shown that by introducing a minimum distance in the coordinate space, the action distributions aquire the energy dispersion specific to the quantum objects. At finite temperature, probability density waves propagating with the sound velocity are obtained as nonstationary solutions of the classical Fokker-Planck equation. The results suggest that in a system of quantum Brownian particles, a transition from complex to real probability waves could be observed.
\\
$\underline{~~~~~~~~~~~~~~~~~~~~~~~~~~~~~~~~~~~~~~~~~~~~~~~~~~~~~~~~
~~~~~~~~~~~~~~~~~~~~~~~~~~~~~~~~~~~~~~~~~~~}$ \\
{\bf PACS:} 03.65.Yz, 05.40.-a, 45.20.Jj \\ 
\newpage
\section{ Introduction}  
Probability waves, beside classical degrees of freedom, describe the outcome of the measurement process in quantum mechanics, or mixed classical-quantum systems. However, well beyond atomic scale, quantum coherence phenomena appear in macroscopic superconducting devices, while many nuclear collective states are related to classical rotational or vibrational degrees of freedom. This wide domain of interplay between classical and quantal indicates that probability waves could be relevant not only for the atomic systems, but also at classical level.  
 \\ \indent
The early attempts to understand atomic phenomena relied on classical analogies, inferring for instance that spectra of hydrogen resemble acoustic phenomena in organ pipes \cite{her}. Though, it was the emission, absorbtion and thermalization of electromagnetic radiation that provided the interface used to shape the general concept of wave-particle duality. 
\\  \indent
In the quantum theory, unlike acoustics, the linearity expressed by the superposition principle holds only for complex wave functions having a nonlinear dependence on the observables, such as probability density. Still, the underlying classical Hamilton equations and the time-dependent Schr\"odinger equation (TDSE) share a common mathematical framework, as both can be obtained from variational principles with suitable action integrals \cite{ks}. This formal aspect allows to derive the mixed classical-quantum dynamics from extended variational principles \cite{pha, hr, cjp}. In particular, a quantum particle coupled to a classical environment at finite temperature can be described using Schr\"odinger-Langevin  \cite{pha, hat} or quantum Fokker-Planck \cite{cjp} equations. Within this approach, the  probability distribution for the quantum component of a mixed system is thermalized by the classical noise, but the quantum backreaction has no such effect on the classical component, in the sense that no random terms appear in the classical equations of motion as a result of purely quantum fluctuations. 
\\ \indent
In semiclassical gravity theory, the metric fluctuations appear as random forces with finite correlation time \cite{ford}. Quantum fluctuations can also be included in time-dependent mean-field dynamics, by assigning to the trial state an effective temperature related to its energy dispersion, and a "quantum Langevin force" \cite{or}. Though, such procedures remain ambiguous, as the various attempts \cite{bm} to understand the transition between classical and quantum fluctuations are incomplete and faced with difficulties. 
\\ \indent
In this work the probability waves are associated with coherent solutions of the classical transport equations, rather than with the continuous action of random forces. The property of coherence is attributed to solutions which evolve  without changing their functional dependence on coordinates and momenta. The kinematics of pure distributions, appearing in classical Hamiltonian systems, is discussed in Sect. 2. An important example is provided by the action distributions, localized in the momentum space. It is shown that by introducing a minimum distance in the coordinate space, the action distributions take the form of the Wigner functions. At finite temperature, the thermal noise can change the statistical ensemble of a pure distribution. This aspect is presented in Sect. 3, considering both stationary and nonstationary solutions of the classical Fokker-Planck equation. The results show that beside the action and quantum waves, relevant at zero temperature, at thermal equilibrium probability waves propagating with the sound velocity can appear. Conclusions are summarized in Sect. 4.
\section{ Classical coherent states } 
The statistical properties of classical systems composed of $N$ particles can be described by a time-dependent distribution function $f \ge 0$ defined on the one-particle momentum phase-space $M$. If $\delta \Omega_m$ is a volume element around the point $m \in M$, then $f(m,t) \delta \Omega_m$ is proportional to the probability to find a particle localized in $\delta \Omega_m$. Therefore, $f$ is normalized by the integrality condition    
\begin{equation}
\int d \Omega_m f(m,t) = N~~,~~N \geq 1~~.
\end{equation} 
Let us consider a one-dimensional system consisting of a single particle ($N=1$) with the Hamiltonian $H(x,p,t)$ depending on the coordinate $x$, the canonical momentum $p$ and time. If there are no external forces, the equations of motion are
\begin{equation}
d_t x= \partial_p H ~~,~~ d_t p = - \partial_x H~~,
\end{equation}
where $d_t \equiv d/dt$ denotes the total derivative with respect to the time $t$, and $\partial_X \equiv \partial / \partial X$ the partial derivative with respect to the variable $X$. \\ \indent
The distribution function $f(x,p,t)$ evolves according to the Liouville equation
\begin{equation}
\partial_t f + {\cal L}_H f =0 \label{lieq}
\end{equation}
where ${\cal L}_H f \equiv - \{ H , f \} $ is the Lie derivative defined by the Poisson bracket \cite{am}
\begin{equation}
{\cal L}_H = (\partial_p H) \partial_x - (\partial_x H) \partial_p  ~~.
\end{equation}
For a Hamiltonian of the form $H(x,p)= p^2/2m + V(x)$, (\ref{lieq}) becomes
\begin{equation}
\partial_t f  + \frac{p}{m} \partial_x f - V' \partial_p f=0 ~~, \label{leq0}
\end{equation}
where $V' \equiv \partial_x V$. To find coherent solutions of this equation it is convenient to use the Fourier transform $\tilde{f}(x,k,t)$ in the momentum coordinate,
\begin{equation}
\tilde{f}(x,k,t) \equiv \int dp ~e^{i kp} f(x,p,t) ~~. \label{fk}
\end{equation}
Thus, if $f(x,p,t)$ is a solution of (\ref{leq0}) then its Fourier transform $\tilde{f}(x,k,t)$ should satisfy 
\begin{equation}
\partial_t \tilde{f} - \frac{i}{m}  \partial^2_{kx} \tilde{f} + i k V' \tilde{f} =0~~. \label{fle}
\end{equation}  
Various local quantities of interest, such as densities of localization probability $n(x,t)$, current $j(x,t)$ and kinetic energy $\epsilon(x,t)$, can be expressed directly in terms of $\tilde{f}$ and its derivatives at $k= 0$ by
\begin{equation}
n(x,t) \equiv \int dp ~ f (x,p,t)  = \tilde{f} (x,0,t)~~,  \label{n}
\end{equation}
\begin{equation}
j (x,t) \equiv  \int dp ~ \frac{p}{m} f(x,p,t)  = - \frac{i}{m} \partial_k \tilde{f}(x,0,t) ~~, \label{j}
\end{equation}
\begin{equation}
\epsilon (x,t) \equiv  \int dp ~\frac{p^2}{2 m} f(x,p,t) = - \frac{1}{2m} \partial^2_k \tilde{f}(x,0,t) ~~. \label{en} 
\end{equation}
In general, $f(x,p,t)$ is specified by the infinite series of partial derivatives
$\partial^\mu_k \tilde{f}\vert_{k=0}$, $\mu =0,1,2,...$. Though, certain coherent solution can be defined only in terms of $n(x,t)$, or a simple functional of $n(x,t)$. Such functionals which satisfy the superposition principle and generate a linear space will be refered as "probability waves".   
\\ \indent
An important class of coherent states for the Liouville equation (\ref{leq0}) are the "action distributions"  
\begin{equation}
f_0 (x,p,t) = n (x,t) \delta( p- \partial_x S(x,t) )~~, \label{cs1}
\end{equation}
which remain all the time a product between $n (x,t)$ and $\delta( p- \partial_x S(x,t) )$. The two real functions of coordinate and time, $n (x,t)$ and $S(x,t)$, are related to the Hamiltonian flow. By Fourier transform (\ref{cs1}) becomes 
\begin{equation}
\tilde{f}_0(x,k,t) = n(x,t) e^{i k \partial_x S(x,t)}   \label{f0}
\end{equation}
while (\ref{fle}) reduces to the system of equations 
\begin{equation}
\partial_t n = - \partial_x j \label{co0} 
\end{equation}
\begin{equation}
n \partial_x [ \partial_t S + \frac{(\partial_x S)^2}{2m} +V]=0 \label{hj}
\end{equation}
where $j \equiv n \partial_x S /m$ denotes the current density (\ref{j}). Thus, by assuming the existence of a "momentum potential" $S(x,t)$, we obtain both the continuity equation (\ref{co0}) and the Hamilton-Jacobi equation \cite{mar} in the form (\ref{hj}). These coupled equations describe "action waves" $n (x,t)$ and can also be derived by using a variational approach (Appendix 1). 
\\ \indent
The partial derivative $\partial_x S(x,t)$ in (\ref{f0}) is the limit of $[S(x+ \ell/2,t)-S(x-\ell/2,t)]/\ell$ when $\ell \rightarrow 0$. If a new parameter $\sigma = \ell / k$ is introduced, then    
\begin{equation}
\tilde{f}_0(x,k,t) = \lim_{\sigma \rightarrow 0} 
\psi^*(x - \frac{\sigma k}{2},t) \psi (x + \frac{\sigma k}{2},t) \label{lim0}
\end{equation}
where $\psi(x,t)$ denotes the complex functional $\psi = \sqrt{n} \exp( i S / \sigma )$\footnote{If $S$ is not single-valued, $\psi$ is a superposition of terms from different branches \cite{jbk}.}. Presuming that, as might be the case in numerical calculations, when both $\ell$ and $k$ decrease to zero the ratio $\sigma=\ell/k$ remains finite, 
\begin{equation}
 \lim_{\ell, k \rightarrow 0} \sigma  >0 ~~,
\end{equation} 
then beside of the limit (\ref{lim0}) we may consider also the "quantum distribution"   
\begin{equation}
\tilde{f}_\psi (x,k,t)  \equiv \psi^*(x - \frac{\sigma k}{2},t) \psi (x + \frac{\sigma k}{2},t) = (\hat{U}_{-k} \psi^*)(  \hat{U}_k \psi)    \label{fq}
\end{equation} 
as a possible coherent solution of (\ref{fle}), presenting its own interest. Here $\hat{U}_k = \exp (\sigma k \partial_x/2)$, while $\sigma$ is a constant. Worth noting, if $\sigma = \hbar$, then $f_\psi$ obtained by inverting (\ref{fk}), 
\begin{equation}
f_\psi(x,p,t)= \frac{1}{2 \pi} \int dk ~e^{-i kp} \tilde{f}_\psi(x,k,t)  
\label{ift}
\end{equation}
is the Wigner transform \cite{gut} of $\psi(x,t)$,  while the general expressions  (\ref{n}), (\ref{j}) and (\ref{en}) correspond exactly to the canonical quantization of the momentum and kinetic energy. Moreover, the probability density  $\nu (p,t)$ of the momentum, 
$$
\nu (p,t) \equiv \int dx ~f_\psi (x,p,t) = \frac{1}{2 \pi} \int dx \int dk ~e^{-ikp} \tilde{f}_\psi (x,k,t)  ~~,   
$$
can be written in the form $\nu(p,t) = \vert \langle \psi_p \vert \psi  \rangle \vert^2$ with 
$$
\psi_p(x)= \frac{1}{\sqrt{ 2 \pi \hbar}} e^{i xp/ \hbar}~~,~~\langle \psi_p \vert \psi \rangle \equiv \int dx ~\psi^*_p(x) \psi(x,t)~~,
$$
showing that (\ref{fq}) also contains the interpretation of the scalar product $\langle \psi_p \vert \psi \rangle$ in the Hilbert space ${\cal H}$ generated by $\psi$ as a probability amplitude. 
 \\ \indent
The transition from the action distribution $\tilde{f}_0$ to the quantum distribution $\tilde{f}_\psi$ could be related to the existence of a limit speed $c$ and a minimum distance $\ell \sim m^{-1}$. Thus, if the limit speed is taken into account by restricting the integral (\ref{fk}) over $p$ to the finite interval $[-mc,mc]$, then (\ref{ift}) becomes a Fourier series. This series is a sum over an infinite set of discrete values of $k$ separated by $1/mc$, and therefore $\sigma \sim mc \ell$ is finite.
\\ \indent 
In the case of $\tilde{f}_\psi$ the first two partial derivatives in (\ref{fle}) are
\begin{equation}
\partial_t \tilde{f}_\psi  = (\hat{U}_{-k} \psi^*)(  \hat{U}_k \partial_t \psi) +
(\hat{U}_{-k} \partial_t \psi^*)(  \hat{U}_k \psi)  
\label{d1}
\end{equation}  
and
\begin{equation}
\partial_{kx}^2 \tilde{f}_\psi = \frac{\sigma}{2}[  
(\hat{U}_{-k} \psi^*)(  \hat{U}_k  \partial_x^2 \psi) - (\hat{U}_{-k} \partial_x^2 \psi^*)(  \hat{U}_k \psi)]~~. \label{d2}
\end{equation}   
To recover an important property of the Wigner transform in the correspondence  between the Liouville equation and TDSE, let us assume that $V'''=0$, so that $k V' \tilde{f}_\psi$ in (\ref{fle}) can be written in the form
\begin{equation}
k V' \tilde{f}_\psi = \frac{1}{\sigma}[  
(\hat{U}_{-k} \psi^*)(  \hat{U}_k V \psi) - (\hat{U}_{-k}V \psi^*)(  \hat{U}_k \psi)]~~. \label{d3}
\end{equation}   
Replacing (\ref{d1}), (\ref{d2}) and (\ref{d3}) in (\ref{fle}) we obtain 
\begin{equation}
(\hat{U}_{-k} \psi^*)( \hat{U}_k \hat{\Lambda} \psi) + (\hat{U}_{-k} \hat{\Lambda}^* \psi^*)(  \hat{U}_k \psi) =0~~, \label{lamb}
\end{equation}  
where $\hat{\Lambda}$ is the linear operator
\begin{equation} 
\hat{\Lambda} \equiv \partial_t - \frac{i \sigma}{2m} \partial_x^2 + \frac{i}{\sigma} V~~.
\end{equation}
Therefore, $\tilde{f}_\psi$ defined by (\ref{fq}) is a solution of (\ref{fle}) if
$\hat{\Lambda} \psi =0$, or 
\begin{equation} 
i \sigma \partial_t \psi = \hat{H} \psi~~,~~
\hat{H}=  - \frac{\sigma^2}{2m} \partial_x^2 + V ~~, \label{se}
\end{equation}
formally the same as TDSE for the complex wave-function $\psi$. A similar result can be obtained in the case of a charged particle in uniform magnetic field, presented in Appendix 2. \\
\indent    
For the harmonic oscillator potential $V(x) = m \omega^2 x^2/2$, (\ref{se}) has both stationary ($\hat{H} \psi_n = E_n \psi_n$) and nonstationary solutions, known as Glauber coherent states
\begin{equation}
\psi^c (x,t)  = ( \frac{\alpha}{ \pi} )^{\frac{1}{4}} e^{- i \omega t /2}
 e^{- \alpha (x- u)^2/2  +  i \alpha v (x- u/ 2)/ m \omega }~~,
\label{gcs}
\end{equation}
where $\alpha$ is a constant while $u$ and $v$ satisfy the equations of motion 
\begin{equation}
 d_t u= \frac{v}{m},~~ d_t v =  - m \omega u ~~.
\label{fv9p}
\end{equation}
By the Wigner transform (\ref{ift}), $\psi^c$  yields the classical distribution
\begin{equation}
f_{\psi^c} (x,p,t) = \frac{ \alpha}{\pi m \omega} e^{ - \alpha( x-u)^2 - \alpha (p-v)^2/ (m \omega)^2} ~~,   \label{gdf}
\end{equation}
representing a Gaussian with a fixed width in the phase-space, oscillating along a classical trajectory.  \\ \indent 
The limitation to polynomial potentials of degree at most $2$ (square well, uniform field, harmonic oscillator), presumed above to derive (\ref{se}), reflects the van Hove theorem on the validity of the canonical quantization \cite{neh}. In general, if $V''' \ne 0$ and $\psi(x,t)$ is a solution of (\ref{se}), then $f_\psi (x,p,t)$ satisfies a modified Liouville equation \cite{bg}. For instance, if $V(x)$ is a quartic polynomial, the modified equation has the form \cite{cv}
\begin{equation}
\partial_t f_\psi  + \frac{p}{m} \partial_x f_\psi - V' \partial_p f_\psi = - \frac{ \sigma^2}{24} V''' \partial_p^3 f_\psi~~. \label{fq3}
\end{equation}
One should note though that by considering $\tilde{f}_\psi (x,k,t) $ as a "matrix" element $\hat{\rho}_{ab} \equiv \psi (x_a) \psi^* (x_b)$ of the density operator $\hat{\rho}$ between $x_a=x+ \sigma k/2$ and $x_b=x- \sigma k/2$, then for $k \approx 0$, $V' =  (V(x_a)-V(x_b)) /(x_a-x_b)$, $ \sigma k V' \tilde{f}_\psi = [ V, \hat{\rho}]_{ab}$ and (\ref{fle}) takes the form of the quantum Liouville equation $i \sigma \partial_t \hat{\rho}_{ab} = [ \hat{H}, \hat{\rho} ]_{ab}$.  
\section{ The thermal sound}  
The equations of motion for a nonrelativistic Brownian particle are
\begin{equation}
d_t x= \frac{p}{m}~~,~~   d_t p = - V' + \xi(t) - \gamma \frac{p}{m} \label{bp}
\end{equation}
where $\gamma$ is the friction coefficient and $\xi$, $- \gamma p/m$, denote the external force, respectively the backreaction produced by the thermal environment. 
\\ \indent
In the bilinear coupling model \cite{zw} $\xi$ is a function $\xi({\cal E}_0,t)$ of time and the variables ${\cal E}_0$ describing the microscopic structure of the environment at the initial moment $t=0$.  If ${\cal F}({\cal E},T)$ is the distribution function of the environment in thermal equilibrium at the temperature $T$, then
\begin{equation}
<< \xi(t) >> \equiv \int d \Omega_{{\cal E}_0} {\cal F}({\cal E}_0,T) \xi({\cal E}_0,t) =0
\end{equation}  
and
\begin{equation}
<< \xi(t) \xi(t') >> \equiv \int d \Omega_{{\cal E}_0} {\cal F}({\cal E}_0,T) \xi({\cal E}_0,t) \xi({\cal E}_0,t') = k_B T \Gamma (t-t')~~, \label{fdt}
\end{equation}  
where $\Gamma (t) $ is the memory function. The phase-space average $<<...>>$ is presumed to be the same as the average $< ..>$ over an ensemble of independent Brownian trajectories. Thus, the  statistical significance of the distribution ${\cal F}({\cal E},T)$ appears indirectly, by the correlation function (\ref{fdt}). For the Brownian particle described by (\ref{bp}), this is specified by $\Gamma (t) = 2 \gamma \delta(t)$. \\ \indent 
Because of friction, a volume element $\delta \Omega \equiv \delta x \delta p$ shrinks according to $d_t \delta \Omega = - \gamma \delta \Omega /m$,  and the particle number conservation $d_t( f \delta \Omega) =0$ implies $d_t f= \gamma f /m$. The definition of the total derivative $d_t f$ 
\begin{equation}
\frac{df}{dt} \equiv  \partial_t f + (d_t x) \partial_xf +(d_tp) \partial_p f ~~,
\end{equation}
and (\ref{bp}) show that the local dynamics of $f$ is expressed by  
\begin{equation}
\partial_t f + \frac{p}{m} \partial_x f - (V' - \xi + \gamma \frac{p}{m})  \partial_p f = \frac{\gamma}{m} f~~, \label{lbr}
\end{equation}  
or
\begin{equation}
\partial_t f =- \frac{p}{m} \partial_x f + \partial_p (V' - \xi + \gamma  \frac{p}{m})f ~~. \label{mle}
\end{equation}  
Due to the noise, the statistical ensemble describing the particle is altered, so that the actual distribution function is the average $<f> (x,p,t) $ of  $f (x,p,t)$ at each phase-space point $(x,p)$. The transport equation satisfied by $<f>$, obtained from (\ref{mle}) by average over an ensemble of solutions $f (x,p,t)$, is the classical Fokker-Planck equation
\begin{equation}
\partial_t <f> + \frac{p}{m} \partial_x <f> - \partial_p ( V' + \gamma  \frac{p}{m})<f> = \gamma k_B T \partial_p^2 <f>  ~. \label{fpl}
\end{equation}  
When describes an environmental particle, its outcome $<f>$ should reproduce self-consistently ${\cal F}({\cal E},T)$. At zero temperature, (\ref{fpl}) admits solutions of the form (\ref{cs1}), with $n(x,t)$ and $S(x,t)$ provided by (\ref{co0}), respectively the modified Hamilton-Jacobi equation 
\begin{equation}
 \partial_t S + \frac{(\partial_x S)^2}{2m} +V+ \frac{\gamma}{m} S=0~~, \label{mhj}
\end{equation}
associated to the classical equations of motion with linear friction \cite{raz}. \\ \indent 
At finite temperature it is convenient to use the Fourier transform 
\begin{equation}
<\tilde{f}>(x,k,t) \equiv \int dp ~e^{i kp} <f>(x,p,t)  ~~,
\end{equation}
and the corresponding transport equation 
\begin{equation}
\partial_t <\tilde{f}> - \frac{i}{m} \partial^2_{xk} <\tilde{f}> + k (i V' +   \frac{\gamma}{m} \partial_k) <\tilde{f}>= - \gamma k_B T k^2 <\tilde{f}>  ~~. \label{fpl1}
\end{equation}  
A series expansion of $<\tilde{f}>(x,k,t)$ at $k=0$,  
$$
<\tilde{f}>(x,k,t) = n(x,t)+ im j(x,t) k 
$$
\begin{equation}
-  2m \epsilon(x,t) \frac{k^2}{2!} 
-i (2 m)^2 \chi(x,t) \frac{k^3}{3!}+ ... ~~ \label{tex}
\end{equation}
provides the ensemble averages of probability (\ref{n}), current (\ref{j}) and kinetic energy (\ref{en}) densities, denoted for simplicity in (\ref{tex}) also by $n$, $j$ and $\epsilon$. When (\ref{tex}) is replaced in (\ref{fpl1}), the consecutive terms of the expansion yield the system of coupled equations 
\begin{equation}
\partial_t n = - \partial_x j  \label{ce}
\end{equation}
\begin{equation}
m \partial_t j = - V' n - \gamma j - 2 \partial_x \epsilon    \label{dj}
\end{equation}
\begin{equation}
\partial_t \epsilon = \frac{ \gamma k_BT}{m} n - V' j - \frac{2 \gamma}{m} \epsilon   - 2 \partial_x \chi ~~,~~... \label{de}
\end{equation} 
If the cubic term $\chi(x,t) = \int dp ~p^3 <f> / (2m)^2$ in (\ref{de}) is neglected, then thermal equilibrium ($\partial_t \epsilon =0$) is reached when  
\begin{equation}
\epsilon = \frac{k_BT}{2} n - \frac{m V'}{2 \gamma} j ~~. \label{epse}
\end{equation}
In this case, a stationary state corresponds to $\partial_t j=0$ and 
\begin{equation}
j = - \frac{2}{\gamma} \partial_x \epsilon - \frac{V'}{\gamma} n
=-D \partial_x n -  \frac{V'}{\gamma} n + \frac{m}{\gamma^2} \partial_x (V'j) \label{j1}
\end{equation}
where $D= k_BT/ \gamma$ is the diffusion coefficient\footnote{If $\partial_x \epsilon =0$ too, then $j= - V' n/\gamma$ and according to (\ref{epse}), $\epsilon$ is the sum between a drift component $m j^2/2n$ and a thermal component $n k_BT/2$.}. At strong friction the last term in (\ref{j1}) can be neglected, and (\ref{ce}) becomes the Smoluchowski equation 
\begin{equation}
\partial_t n=  \partial_x (D \partial_x n +  \frac{V'}{\gamma} n)~~.
\end{equation}
In states of nonequilibrium at low temperature, the kinetic equation (\ref{leq0}) provides the velocity of zero sound \cite{thou}. Similarly, applied to a nonstationary state, (\ref{ce}) and (\ref{dj}) yield 
\begin{equation}
\partial^2_t n= - \partial^2_{xt} j =  \frac{2}{m} \partial^2_x \epsilon + \frac{1}{m} \partial_x (V'n) - \frac{\gamma}{m} \partial_t n ~~. \label{nste}  
\end{equation}
If $V=0$ and $\epsilon$ is given by (\ref{epse}), then (\ref{nste}) becomes a wave equation for the probability density,
\begin{equation}
\partial^2_t n= v_s^2 \partial^2_x n  - \frac{\gamma}{m} \partial_t n ~~, 
\label{cpw1}
\end{equation}
where $v_s = \sqrt{k_BT/m} $ is the sound velocity provided by Newton's formula. \\ \indent
It is interesting to remark that if $\tilde{f}_\psi$ of (\ref{fq}) is expanded in the form (\ref{tex}), then $j=n \partial_xS/m $. $\epsilon= mj^2/2n + \sigma^2 [ ( \partial_x \sqrt{n})^2 - \sqrt{n}  \partial_x^2 \sqrt{n}]/4m$, so that when $T=0$, $\gamma=0$, (\ref{ce}) and (\ref{dj}) yield 
\begin{equation}
\partial_t n = - \partial_x j~~,~~  n \partial_x [ \partial_t S + \frac{(\partial_x S)^2}{2m} -  \frac{\sigma^2}{2m} \frac{ \partial_x^2 \sqrt{n} }{ \sqrt{n}} + V]=0 ~~, \label{56}
\end{equation}
close to (\ref{co0}), (\ref{hj}) and equivalent to the TDSE (\ref{se}). Moreover, if $<\tilde{f}_\psi>(x,k,t)$ is expressed as a  matrix element $< \hat{\rho} >_{ab} (t)$ of the average density operator $<\hat{\rho}>$ between $x_a=x+ \sigma k/2$ and $x_b=x- \sigma k/2$, then for $k \approx 0$ (\ref{fpl1}) takes the form of the quantum Fokker-Planck equation \cite{cl}. 
\section{ Summary and conclusions} 
Probability waves in the configuration space are related to certain solutions of the Liouville equation which are coherent, in the sense that they keep in time the same functional dependence on the coordinate and momentum. Of course, this kinematical property alone does not explain how such distributions are formed, but only shows that once created, they are stable. \\ \indent
Two generic examples of classical, functional coherent states, are presented in Sect. 2. The first example concerns the action distributions (\ref{cs1}), localized in the momentum space at the value given by the generating function $S$. For the related action wave the current ($j$) and kinetic energy ($\epsilon$) densities are expressed in terms of only two functions, the probability density $n$ and $S$, provided by the continuity, respectively the Hamilton-Jacobi equations. These equations can also be obtained from a least-action principle in which $n$ and $S$ are canonically conjugate (Appendix 1). \\ \indent
The second example is derived from the first, presuming the existence of a minimum distance in the coordinate space which decreases as a function of the inertial parameter. Worth noting, this elementary distance may not necessarily result from a lattice structure of the phase-space, but take into account a minimum interval of time \cite{cjp}. It is shown that if the space derivative of $S$ in the action distribution is written as finite difference, then one obtains a Wigner-type distribution (\ref{ift}), expressed in terms of a single complex functional $\psi$ of $n$. However, by contrast to the action distributions, the coherence of the Wigner distributions is maintained only by a harmonic oscillator, uniform field, or square well potential. For the harmonic oscillator, beside the static probability distributions can be found Gaussian wave-packets in phase-space (\ref{gdf}), oscillating along the classical orbit. Such solutions could be useful for instance to describe a macroscopic Bose-Einstein condensate in a harmonic trap \cite{kett}. \\ \indent
The evolution of the average phase-space distribution function for a Brownian particle is discussed in Sect. 3. Containing stochastic and dissipative forces, the equations of motion (\ref{bp}) yield the modified Liouville equation (\ref{mle}) and then, by ensemble average, the Fokker-Planck equation (\ref{fpl}). The Fourier transform in momentum of the average distribution function provides the localization probability ($n$), current ($j$), and kinetic energy ($\epsilon$) densities, while the pure distributions discussed in Sect. 2 may still be useful to obtain solutions in the limit of zero temperature. However, in general (\ref{fpl}) describes the irreversible increase of the entropy, and the functional relationship between $n$, $j$ and $\epsilon$ is specified only by the system (\ref{ce}), (\ref{dj}), (\ref{de}). In free space at thermal equilibrium this system reduces to the sound waves equation (\ref{cpw1}), while for a Wigner distribution at $\gamma, T$ zero it provides the Schr\"odinger equation (\ref{56}). As both equations arise in the same framework, the result of thermal averaging in nonrelativistic quantum systems could be a specific form of quasiclassical behaviour, expressed by a transition from complex ($\psi$) to real ($n$) probability waves, as an intermediate stage between decoherence and complete dissipation. One should note though that at strong friction this stage can be suppressed, and in the previous calculations on atomic tunneling at finite temperature \cite{hat}, it was not observed.  
\\ \indent
The assumption $<< \xi(t) \xi(t') >>  \sim \delta(t-t')$ in (\ref{fdt}), of $\delta$-correlated noise,  is reflected by the second-order derivative of the distribution function in the Fokker-Planck equation (\ref{fpl}). This term is essential to retrieve the nonrelativistic wave equation (\ref{cpw1}), but is not suitable to simulate effects of the quantum fluctuations, such as the third-order derivative $\partial_p^3 f_\psi$ in (\ref{fq3}). The emergence of the quantum dynamics out of a discontinuous character of the inertial motion, rather than by the action of random forces, might provide an alternative worth of further consideration. 
\section{Appendix 1: Action functional for action waves } 
The coupled equations (\ref{co0}), (\ref{hj}) describing action waves can also be derived using the variational principle $\delta_{n,S} {\cal A} =0$ for the action functional \cite{hr} 
\begin{equation}
{\cal A}[n,S] = - \int dt \int dx ~n [\partial_t S + \frac{(\partial_x S)^2}{2m} +V] \label{af} 
\end{equation}
with respect to the local variations of the fields $n(x,t)$ and $S(x,t)$. The variational formulation shows that the coupled continuity and Hamilton-Jacobi equations describe an infinite-dimensional Hamiltonian system, having the generating function $S(x,t)$ and the density $n(x,t)$ as conjugate variables. The symplectic form and the Hamilton function of this system are, respectively
\begin{equation}
\omega = \int dx ~(d n \wedge dS)   \label{sf}
\end{equation} 
and 
\begin{equation}
H_{\cal A} = \int dx ~n [\frac{(\partial_x S)^2}{2m} +V] ~~. \label{ha}
\end{equation}
In this framework, the conservation of the particle number expressed by (\ref{co0}) reflects the "gauge-invariance" of the variational equation $\delta_{n,S} {\cal A}=0$ with respect to the change of $S$ by adding an arbitrary function of time.  
\\ \indent
It is interesting to note that (\ref{af}), (\ref{sf}) and (\ref{ha}) define a Hamiltonian system in the Hilbert space ${\cal H}$ generated by $\psi = \sqrt{n} \exp( i S / \sigma )$, while the equality 
\begin{equation}
\int dx ~(d n \wedge dS) = - i \sigma \int dx ~(d \psi^* \wedge d \psi)  
\end{equation} 
proves that $\omega$ coincides with the symplectic form \cite{cm} induced by the complex structure of ${\cal H}$.
\section{Appendix 2: Quantum waves in magnetic field } 
In the three-dimensional space, an electric charge in the magnetic field ${\bf B} = \nabla \times {\bf A}$, is described by $H= ({\bf p} - {\sf q} {\bf A})^2/2m +V$, and (\ref{leq0}), (\ref{fle}) take the form 
\begin{equation}
\partial_t f  + \frac{{\bf p} - {\sf q} {\bf A}}{m} \cdot \nabla f - [\nabla \cdot \nabla_p~, V ]f +  \frac{{\sf q}({\bf p} - {\sf q} {\bf A}) }{m}  \cdot [\nabla \cdot \nabla_p~, {\bf A}]f=0  ~~, \label{cle} 
\end{equation}
respectively
\begin{equation}
\partial_t \tilde{f} - \frac{i}{m}  \nabla \cdot \nabla_k \tilde{f} + i [ {\bf k} \cdot \nabla~ ,  V + \frac{{\sf q}^2}{2m} {\bf A}^2] \tilde{f} 
- \frac{\sf q}{m} ({\bf A} \cdot \nabla + \nabla_k \cdot [ {\bf k} \cdot \nabla~ , {\bf A}] ) \tilde{f} =0~~, \label{a2} 
\end{equation}  
where $\nabla \equiv \partial / \partial {\bf r}$, $\nabla_p \equiv \partial / \partial {\bf p}$ and $\nabla_k \equiv \partial / \partial {\bf k}$. 
\\ \indent
For a quantum nonrelativistic scalar particle the coherent solutions of (\ref{a2}) have the form 
\begin{equation}
\tilde{f}_\psi ({\bf r},{\bf k},t)  \equiv \psi^*( {\bf r} - \frac{\hbar {\bf k}}{2},t) \psi ({\bf r} + \frac{\hbar {\bf k}}{2},t) = (\hat{U}^{-1} \psi^*)(  \hat{U} \psi)    
\end{equation} 
with $\hat{U} = e^{\hbar {\bf k} \cdot \nabla/2}$. Presuming that ${\bf A} = ({\bf B} \times {\bf r})/2$, with ${\bf B}$ constant, and $({\bf k} \cdot \nabla)^3 V =0$, then for the first three terms of (\ref{a2}) we obtain 
\begin{equation}
 \partial_t \tilde{f}_\psi - \frac{i}{m}  \nabla \cdot \nabla_k \tilde{f}_\psi + 
i [{\bf k} \cdot \nabla , V + \frac{{\sf q}^2}{2m} {\bf A}^2 ] \tilde{f}_\psi =  \label{lam1}
\end{equation}
$$
(\hat{U}^{-1} \psi^*) (\hat{U} \hat{\Lambda}_1 \psi) + (\hat{U}^{-1} \hat{\Lambda}_1^* \psi^*) (\hat{U}  \psi )  
$$
where 
\begin{equation} 
\hat{\Lambda}_1 \equiv \partial_t - \frac{i \hbar^2}{2m} \nabla^2 + \frac{i}{\hbar} (V+ \frac{{\sf q}^2}{2m} {\bf A}^2) ~~.
\end{equation}
\\ \indent
Because  $\nabla \cdot {\bf A}=0$, $\nabla_k \cdot [ {\bf k} \cdot \nabla , {\bf A}]=  [ {\bf k} \cdot \nabla , {\bf A}] \cdot \nabla_k$, and in the last term of (\ref{a2}) 
\begin{equation}
({\bf A} \cdot \nabla + [ {\bf k} \cdot \nabla , {\bf A}] \cdot \nabla_k  ) 
\hat{U} \psi =  \hat{U} ({\bf A} \cdot \nabla) \psi~~. \label{an}
\end{equation}
Thus, with (\ref{lam1}) and (\ref{an}), (\ref{a2}) takes the form
\begin{equation}
(\hat{U}^{-1} \psi^*)( \hat{U} \hat{\Lambda}_2 \psi) + (\hat{U}^{-1} \hat{\Lambda}^*_2 \psi^*)(  \hat{U} \psi) =0~~,
\end{equation}  
where $\hat{\Lambda}_2 = \hat{\Lambda}_1 -{\sf q} {\bf A} \cdot \nabla /m$. This equation reduces to $\hat{\Lambda}_2 \psi =0$, or $i \hbar \partial_t \psi = \hat{H} \psi$, with
\begin{equation} 
\hat{H} =   \frac{(- i \hbar \nabla -{\sf q} {\bf A})^2}{2m} + V ~~. \label{h0}
\end{equation}
\\ \indent
For a wave-function $\Psi$ with two components, describing a particle with spin 1/2, (\ref{a2}) may have coherent solutions of the form 
\begin{equation}
\tilde{f}_\psi ({\bf r},{\bf k},t)  \equiv (\hat{U}^{-1}_s \Psi)^\dagger ( \hat{U}_s \Psi)    
\end{equation} 
where $\hat{U}_s = e^{2 ({\bf k} \cdot {\bf \hat{s}}) (\nabla \cdot {\bf \hat{s}}) / \hbar}$, and ${\bf \hat{s}}$ is the spin operator. With $\hat{U}_s$ instead of $\hat{U}$, to the first order in ${\bf k}$ (\ref{lam1}) remains the same, but (\ref{an}) is changed to
\begin{equation}
({\bf A} \cdot \nabla + [ {\bf k} \cdot \nabla , {\bf A}] \cdot \nabla_k  ) \hat{U}_s \Psi =  \hat{U}_s ( {\bf A} \cdot \nabla  + \frac{i}{2 \hbar} {\bf \hat{s}} \cdot {\bf B}) \Psi~~,
\end{equation}
so that (\ref{h0}) becomes
\begin{equation} 
\hat{H}_s = \hat{H} - \frac{\sf q}{2 m} {\bf \hat{s}} \cdot {\bf B} ~~. \label{hs} 
\end{equation}
Therefore, the solution $\Psi({\bf r},t)$ of the TDSE $i \hbar \partial_t \Psi = \hat{H}_s \Psi$ for a particle with spin corresponds to a coherent solution 
\begin{equation}
f_\Psi({\bf r},{\bf p},t)= \frac{1}{(2 \pi)^3} 
\int d^3k ~ e^{-i {\bf k} \cdot {\bf p} } (\hat{U}^{-1}_s \Psi)^\dagger (  \hat{U}_s \Psi)
\end{equation}
of the classical Liouville equation (\ref{cle}). One should note though that in (\ref{hs}) the magnetic moment ${\sf q} \hbar / 4m$ is half the value provided by the relativistic Dirac equation, and a complete description of a charged particle with spin should include the internal degrees of freedom also in the Liouville equation.

\end{document}